# Wavelength-independent performance of femtosecond laser dielectric ablation spanning over three octaves

Mario Garcia-Lechuga [1,2,*], Oliver Utéza [1], Nicolas Sanner [1], David Grojo [1†]

[1] *Aix Marseille Université, CNRS, LP3, UMR7341, 13288 Marseille, France*
[2] *Instituto de Optica Daza de Valdes (IO), CSIC, Serrano 121, 28006 Madrid, Spain*

Ultrafast laser breakdown of wide bandgap dielectrics is today a key for major technologies ranging from 3D material processing in optical materials to nanosurgery. However, a contradiction persists between the strongly nonlinear character of energy absorption and the robustness of processes to the changes of the bandgap/wavelength ratio depending on applications. While various materials and bandgaps have been studied, we concentrate here the investigations on the spectral domain with experiments performed with wavelength drivers varied from deep-ultraviolet (258 nm) to mid-infrared (3.5 µm). The measured fluence thresholds for single shot ablation in dielectrics using 200-fs pulses exhibit a plateau extending from the visible domain up to 3.5-µm wavelength. This is accompanied, after ablation crater analysis, by a remarkable invariance of the observed ablation precision and efficiency. Only at the shortest tested wavelength of 258 nm, a twofold decrease of the ablation threshold and significant changes of the machining depths are detected. This defines a lower spectral limit of the wavelength-independence of the ablation process. By comparison with simulations, avalanche ionization coefficients are extracted and compared with those predicted with the Drude model. This must be beneficial to improve predictive models and process engineering developments exploiting the new high-power ultrafast laser technologies emitting in various spectral domains.

* mario.garcia.lechuga@csic.es
† david.grojo@univ-amu.fr

## INTRODUCTION

A unique feature with the use of tightly focused ultrashort laser pulses is the ability to achieve localized energy deposition, down to the sub-micrometer scale at the surface or in the bulk of any materials, even those that are initially transparent to the laser wavelength [1]. This makes femtosecond (fs) lasers a key tool for inducing and controlling transformations of various materials with applications ranging from ultraprecision fabrication of fluidic or optic microdevices [2,3], nanosurgery [4] to warm dense matter physics [5].

Over decades, the efforts on both experimental and theoretical researches [6] have allowed to draw a general picture of ultrafast laser dielectric excitation and the resulting potentially extreme conditions accessible inside the matter. As reviewed in detail by Balling and Schou [7], and briefly summarized here, the process is triggered by the generation of free carriers induced by strong-field nonlinear ionization (SFI) mechanisms, as multiphoton (MPI) and tunneling ionization (TI), and subsequently increased by impact ionization (IMP). If the excitation is high enough, the transferred energy from the free electron sub-system to the lattice induces high local temperature and pressure conditions, responsible for material transformations. If the irradiation is performed at the surface, material expansion and ablation can occur [8,9]. This is the basis of industrial micro-machining [10] process solutions or high-precision surgery applications [4,11].

The constantly growing interest on this flexible technology is today accompanied by the advent of new source technologies, opening an enormous palette of accessible parameters, as pulse duration, repetition rate or wavelength that are all supposed to influence the strongly nonlinear material responses and so the experimental outcomes. Therefore, the influence of each parameter remains a question of topical interest for fundamental and industrial perspectives.

On one hand, the role of the repetition rate and pulse duration has been widely explored. For example, studies focusing on repetition rate influence have found benefits on the optimization of waveguide writing at MHz repetition rates [12] and ablation efficiency and cleanliness by using ultrafast pulse trains at GHz repetition rate [13]. Experiments focusing on the pulse duration have revealed that conventional longer femtosecond pulses (> 100 fs) leads to significant IMP contributions responsible for plasma absorption assisting laser energy deposition [14–16], whereas energy deposition with the shortest pulses takes more profit from MPI [17,18] and even TI when few-cycles pulses are applied [19,20].

On the other hand, the wavelength has been the subject of less investigations. Most probably this is because the fundamental wavelength of most available ultrafast laser technologies remained confined until recently in a relatively narrow spectral domain in the near infrared (typ. 800 nm or 1.03 µm for resp. Ti:Sapphire or Yb-doped-crystals technologies). With the recent advent of new compact technologies delivering increased power in new spectral regions and in particular the mid-infrared domain (incl. fiber technologies), one should reasonably expect the possibility to cause drastic changes in the balance between the above described physical processes. By affecting the relative role of MPI, TI and IMP, one can look for optimization on the drivers for energy deposition. This introduces the timeliness for a debate of the achievable performance for precision processing considering the full spectrum covered with the newly available sources.

From a fundamental perspective, one can refer to the Keldysh theories that directly account for the wavelength dependence of nonlinear ionization [21,22] For the experimental works (including ours), one usually relies on studies with optical parametric amplifiers (OPA). These become routinely available in laboratories and allow a continuous tuning of radiations for assessing the detailed spectral dependencies of material responses.

The literature on this subject remains scarce but it is important to refer to the pioneering works of Simanovskii et al. [23] and Jia et al. [24], measuring the breakdown fluence thresholds at the surface of several dielectrics with varying wavelength conditions. Despite the gap separating the spectral domains which were investigated in these two works, one can already extrapolate from their combination a relative invariance of the apparent nonlinear absorption rates (within experimental uncertainty) in a broad spectral region. It is only at the tested spectral limits that a significant change could be observed. In their study concentrating in the mid-infrared, Simanovskii et al [23] measured a drop for the breakdown threshold fluence when increasing the wavelength above 5µm. According to the electronic transition picture described above, this was attributed to a breakdown process that becomes seeded by TI, a wavelength-independent process [25] but driven by IMP with efficiency scaling with the wavelength according to a simple Drude model. However, the investigated spectral domain up to 7.8 µm is at the edge of the vibrational absorption bands, an aspect that raises question on an interpretation solely based on the ionization yield. A similar increase of absorption efficiency is also reported approaching the UV part of the spectrum by Jia et al. [24]. While this can be directly interpreted by increased ionisation rates with a pure MPI picture, the authors had also to introduce significant contributions from sub-conduction band transitions to quantitatively describe the measured process efficiency at short wavelengths.

It is only recently that more experimental investigations have been added on these questions. Gallais and coauthors [26] measured femtosecond laser induced damage thresholds of various bandgap materials (1 to 10 eV) in a spectral region from 310 to 1030 nm. While the paper was concentrating on thin-film materials, the measured thresholds for silica substrates and the reported wavelength dependence compares reasonably well with Jia et al. [24]. For recent investigations turned toward the mid-infrared part of the spectrum, one can turn to Migal *et al.* [27] who investigated the breakdown response of $SiO_2$ and $MgF_2$. The reported thresholds were strongly deviating from those in the pioneer's paper of Simanovskii *et al.* [23]. However, this and another work conducted by Austin et al. [28] on narrow gap material as Germanium concluded that the conventional nonlinear ionization models fail in describing the mid-infrared response. Multi-band and/or intra-conduction band effects were then considered to discuss the obtained results. Among the other related works, it has been studied the laser wavelength influence on the direct writing of optical waveguides or other buried structures in glasses [29] and in silicon [30]. Additionally in silicon surface, we reported the optimization of amorphization by tuning the laser wavelength (258-nm to 4-µm) [31] and Otobe theoretically studied the wavelength dependence (800-nm to 3-µm) of laser-excitation [32]. We can also mention the study of laser-induced breakdown in water [33] or on ablation of corneal stroma [11]. While all are relevant for fundamental and technological considerations, the results obtained in these specific configurations are more hardly comparable to the previously mentioned papers focused on material dielectric ablation.

In this paper, we bridge the gap between these previous reports by investigating the femtosecond laser ablation response in an unprecedented spectral range from 258 nm to 3500 nm. We find a nearly-invariant observables including ablation fluence thresholds, thermally affected zones or maximum crater depths on a very large spectral domain from the visible to the mid-infrared part of the spectrum. Supported by simulations, we concentrate the analyses on sapphire and fused silica, dielectric materials of reference both in applications and in the fundamental study of laser-matter interaction. Simple nonlinear absorption rate considerations lead to the report of apparent avalanche rates, which are compared with those that can be derived by a Drude model. The gained knowledge on the spectral dependence of femtosecond laser machining is discussed in terms of potential optimizations in model and application developments.

## METHODS

### A. Laser system and optical set-up

The multiwavelength study is performed by using a commercial femtosecond amplifier (Pharos, Light Conversion) emitting at 1030 nm with pulse energies up to 500 µJ and 180±10 fs pulse duration at full-width at half maximum (FWHM), characterized by single-shot autocorrelation (TiPA, Light conversion). Tunable near-infrared (NIR) and mid-infrared (MIR) radiation is accessible through optical parametric amplification (Orpheus, Light Conversion), designed for efficient conversion on a large infrared range (signal: 1.5 μm -2.06 μm, idler: 2.06 – 5 μm). In this work, we performed experiments up to 3.5 μm (MIR), due to the limited pulse energy on longer wavelengths. Indeed, the use of tight focusing conditions is avoided in order to maintain the same *ex-situ* characterization methodology (as explained later on) for all the analyses. Visible (VIS) and ultraviolet (UV) radiation (515 and 258 nm) are obtained through harmonic generation (Hiro, Light Conversion). For all the wavelengths, the beam irradiates the targets at normal incidence with linear polarization.

Using the already mentioned autocorrelator device, we measured pulse durations of 190±10 fs FWHM on the range from 1.55-μm to 2- μm and 160±10 fs FWHM at 515-nm. Temporal characterization on the idler spectral range could not been performed with the employed instrument. However, a direct correspondence of the pulse duration is expected between signal and idler, given the optical parametric amplification process generating both beams together. For the deep ultraviolet, we also expect a pulse duration very close to the one measured at 515-nm according to the design (thickness) of the last doubling crystal for generating the 258-nm radiation.

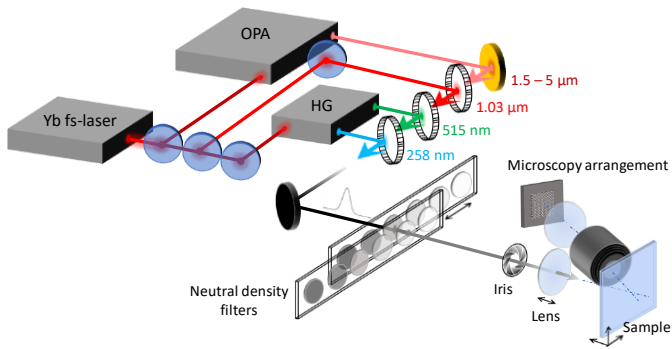

**Figure 1.** Sketch of the experimental set-up. On the top part is represented the beams of different wavelengths obtained from the same master femtosecond laser through harmonic generation (HG) , for ultraviolet and visible radiation, and optical parametric amplification (OPA), for infrared radiation. On the bottom part, it is represented the main optical elements of the irradiation configuration including an *in situ* microscopy system for precise positioning of the focal spot on the surface. The OPA beams are directed toward the irradiation setup using gold mirrors. A selection of dielectric mirrors is used for the other wavelengths (removable mirrors shown by empty volumes).

A sketch of the experimental set-up is shown in Figure 1. Experiments are performed on the basis of single-shot irradiations, by making use of a pulse-picker at the output of femtosecond amplifier. For each selected wavelength, the beam is directed toward the same irradiation beam line. Gold mirrors are used for the beams in the mid-infrared domain covered by the OPA. For the other beams, these are systematically replaced by specific high-reflectivity dielectric mirrors to reduce energy losses and assure excellent spectral filtering at the considered wavelength. The selected beam is then focused normally onto the sample surface by means of an uncoated UV-fused silica aspheric lens (nominal $f = 50$ mm) for the spectral range between 248 nm to 2 μm, and a calcium fluoride plano-convex lens (nominal $f = 25$ mm) from 2.2 μm to 3.5 μm. These choices of materials and focal distances were made because of two reasons: the loss of transparency of UV-fused silica in the infrared range and the reduction of accessible energy in the infrared range. A ring-actuated iris diaphragm is situated as close as possible to the lens to truncate all beams (transmitted power ~75%) and so we always irradiate the samples with well-defined profiles. This is of importance to suppress potential biases in the analyses caused by some of the OPA beam profiles exhibiting pedestals [34].

On the beam path, a motorized wheel with neutral density metallic filters (1 mm UV-fused silica substrate with nickel coating) is placed. The broadband response of the filters allows to control the pulse energy on the full spectral region of interest in this work. For all experiments, the applied power is measured at 1 kHz repetition rate with a thermal powermeter (3A, Ophir), which is calibrated for the spectral range of interest. Peak-to-peak pulse energy stability is measured a pyroelectric energy meter (PE9, Ophir) to be less than 4 % for all the considered wavelengths, being sufficiently low for making unnecessary a measurement of the energy of each applied pulse. Since the powermeter is placed just before the lens, the energy at the sample position is obtained by simply accounting for the Fresnel reflection losses expected on the lens surfaces.

The sample is mounted on a XYZ motorized stages and aligned perpendicular to the irradiation axis. Precise parallelism of the sample with respect to motion axes is achieved by using a 2-axis kinematic sample holder. Motion in the XY plane, perpendicular to the axis of laser incidence, allows to position the sample on a fresh surface before each irradiation. Optimal focusing position is determined empirically by a Z-scanning and surface imaging with an in-situ microscopy (10x microscope objective, tube lens and CCD-camera) placed at 45° as shown in Figure 1. Repositioning of the different samples at the best focal position is guaranteed by exploiting the ≈ 10-μm depth of field of the microscopy system.

### B. Samples and metrology

In this work we study four different samples of three dielectric materials widely used in optics and photonics applications: sapphire, soda-lime glass and fused silica (amorphous, a-SiO$_2$). The sapphire sample is a sapphire window from Meller Optics, Inc. (2 mm thick). The soda-lime glass sample is a microscope slide from Corning (1 mm thick). Two different samples of fused silica were studied. First a sample of synthetic fused silica (Suprasil 2) from Heraeus of 3-mm thickness, being superpolished by Thales-SESO (with measured Ra = 0.2 nm). The same sample was used for some other studies conducted by our group [15]. For this work, it is named as "UV-fused silica". Second is a sample of fused quartz provided by Thuet-France (1 mm thick), named on here as "IR-fused silica". The transmittivity in the ultraviolet region of the spectrum differs on these two samples, being observed by spectrophotometry measurements (UV-2600, Shimadzu, range 190-1500 nm). Measurable absorption is found at 250-nm (~5 eV) for IR-fused silica. In contrary, the full transparency in this UV domain for UV-fused silica is consistent with the expected absorption edge according to the bandgaps considered in the literature from 7.75 to 9.9 eV (7.75 eV used by Gallais et al. [35], 9.6 eV reported by Ravindra et al. [36], 8.9 eV indirect or 9.9 eV direct bandgap reported by Tan et al. [37]). The same measurements performed for the other samples reveal absorption under ~350 nm in soda-lime glass corresponding to an apparent optical bandgap of ~3.5 eV. The absence of measurable absorption in sapphire for the entire measured spectral range is consistent with the band gap of 8.8 eV from the literature [38].

For characterizing the laser induced modifications on the different samples, we use confocal microscopy (Leica DCM3D). Under monochromatic illumination (460 nm) and using a 150x objective lens (NA 0.95) the laser induced modifications are characterized with transverse sub-micrometre spatial resolution and nanometric vertical resolution. This technique allows to characterize the ablated area on different samples under a removal depth-based criterion suppressing any potential subjectivity in the analyses. This is a major advantage in comparison to methods relying on conventional optical microscopy where the interpretation based on an apparent optical contrast which can be dependent on the nature of modifications and material responses. Another advantage of the confocal microscopy method compatible with a well-defined and fixed threshold criterion is the possibility to achieve automated crater analyses using a software-solution (Mountains 8, Digital Surf) as demonstrated in previous works [39,40].

In Figure 2 (a), we show examples of craters produced on each of the mentioned samples, after irradiation by a single pulse at 1.7-µm wavelength. The grey scale of the images is set to show in black the surface depressions and in white the surface elevations. Subsequent analyses of the images allow to obtain the total ablated area (depression region taken under -10-nm level) and the crater profile (see Supplemental figures S1, S2 and S3). .

As part also of the metrology, firstly, each irradiation condition (each energy on each material) was repeated three time in order to add statistics to our analysis. Secondly, the fluence threshold in this work is obtained by following the Liu's method [41]. However, this is rigorously valid for perfect Gaussian beams. This deviated from some of our OPA beam profiles exhibiting a pedestal (e.g. see in ref. [39] the profile at 1.55 µm) and it is the reason of the truncation of all our beams by using a circular aperture (see before). In this way, we do not irradiate with Gaussian spots but systematically with well-defined Airy-like beam shapes that allow to calculate for correction factors to the threshold determination. The details corresponding to this applied procedure extending the Liu's method to imperfect beams is described and demonstrated in ref. [42].

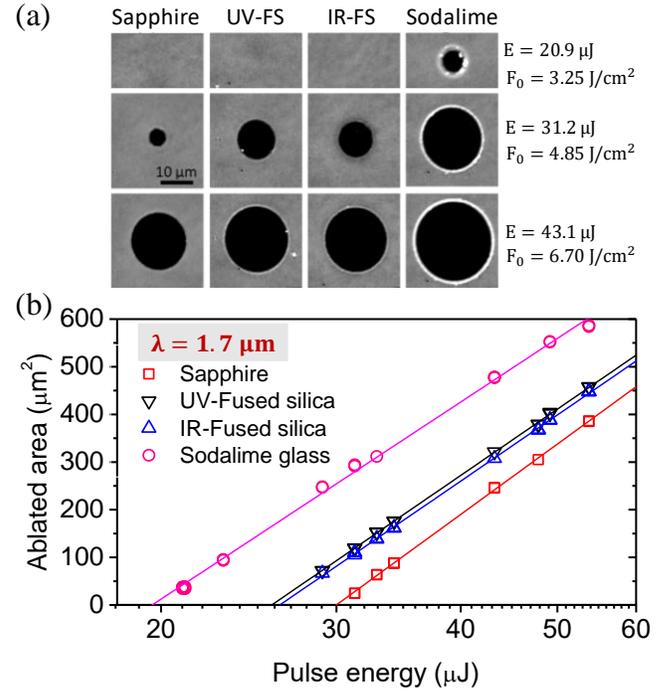

**Figure 2.** (a) Confocal microscopy images (profilometry) of craters induced in four different dielectrics with 1700-nm laser pulses of different energies. The grey scale is set so that depressions and elevations are represented respectively with darker and brighter tones with respect to grey background. (b) Ablated area as a function of the pulse energy for four different materials. Lines correspond to numerical linear regressions following Liu's methodology.

In Figure 2 (b) is shown the representation following the Liu's methodology, where the ablated area of craters shown in Figure 2(a) are represented as a function of the pulse energy (in logarithmic scale). A linear regression enables to extract an energy threshold ($E_{th}$) and an equivalent Gaussian beam waist ($w_{0,Liu}$). The fluence threshold value is then obtained as $F_{th} = (2E_{th}/\pi w_{0,Liu}^2)\eta_{E_{th}}\eta_F$, with $\eta_{E_{th}}$ and $\eta_F$ the corrections factors to apply for a truncated beam (Airy-like beam on target) [42]. Rigorously, those corrections factors depend on the power transfer of the aperture ($P_T$) and the maximum excitation energy considered for the linear fitting procedure (M.E.) [42]. $P_T$, M.E. and values of the retrieved $w_{0,Liu}$ can be found in the Supplemental Note 1.

## C. Modelling

For evaluation of the relative roles of the ionization mechanisms depending on tested radiations, we perform calculations of the temporal evolution of the electron density on the conduction band, $n_e(t)$ according to the standard electron density single rate equation,

$$\frac{dn_e(t)}{dt} = \frac{n_T - n_e}{n_T}\left(W_{SFI}(I(t)) + \alpha \cdot I(t) \cdot n_e(t)\right) \quad \text{Eq.1}$$

where $n_T$ is total density of accessible electrons (valence electron density) in the medium as described by Christensen and Balling [43] and fixed at $n_T = 1 \cdot 10^{23}~cm^{-3}$. The first term, $W_{SFI}$, is the photoionization rate by SFI (MPI or/and TI) and the second term represents the contribution of IMP, where $\alpha$ is the avalanche coefficient and $I(t)$ is the instantaneous laser intensity. For simplification, here we neglect all the relaxation sources (incl. electron trapping mechanisms).

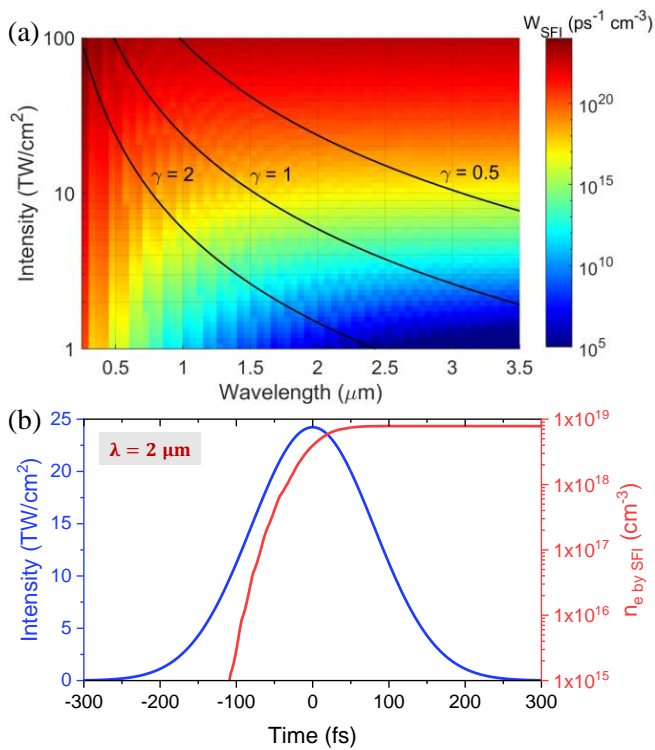

**Figure 3**. (a) Numerical calculation of the strong-field photoionization rate ($W_{SFI}$) as a function of the wavelength and intensity (ranges investigated in this work) following the Keldysh formulation for sapphire (bandgap equal to 8.8 eV) and considering $m^*=m_e$. To assess the transition from TI ($\gamma\ll1$) to MPI ($\gamma\gg1$), contour lines for given adiabatic coefficients ($\gamma = 2, \gamma = 1, \gamma = 0.5$) are appended. (b) (Blue line) Intensity temporal profile of a pulse of 190 fs (FWHM) for a fluence of 4.9 J/cm² corresponding to the measured fluence threshold for ablation of sapphire at $\lambda = 2$ µm. (Red line) Corresponding build-up of the free electron population expected form the strong field ionization (time integral).

The term $W_{SFI}$ is calculated from Keldysh formalism [21], dependent on the conditions of irradiation (laser intensity and wavelength) and on the material properties (material band gap). In this work, we perform calculations for sapphire and UV-fused silica assuming bandgaps of 8.8 eV [38] and 8.9 eV [37] respectively. As an example, the figure 3 (a) shows the calculated $W_{SFI}$ values for sapphire on the range of laser wavelengths and laser intensities investigated in this work. On the same graph, we append several iso-contour lines representing the Keldysh adiabatic parameter $\gamma = \omega\sqrt{m^* \cdot BG}/(e \cdot E)$, ($m^*$, $E$, and $BG$ are respectively the reduced mass of the electron-hole pair, the laser electric field, and the band gap of the material). This parameter is usually referred when discussing the relative importance of MPI and TI. When $\gamma \gg 1$ MPI is expected to dominate, corresponding for the calculated example to irradiations in the UV-visible range or when irradiating on the NIR with intensities well under the modest TW/cm² level. When $\gamma \ll 1$ TI dominates, corresponding to extremely strong laser fields or long wavelengths regimes (MIR) with intensities exceeding 10 TW/cm².

An example of application of equation 1 when accounting only its first term (only SFI) is shown in figure 3 (b). Plotted in blue is represented the temporal profile in intensity for a 190-fs pulse. The integration over time leads to a fluence value of 4.9 J/cm², which, as shown later on, corresponds to the ablation fluence threshold of sapphire at a wavelength of $\lambda = 2$ µm. Additionally, the free electron generation by SFI in sapphire is calculated according to Eq. 1 using these pulse characteristics. Under these conditions, we note that the photoionization channel leads to the creation of a free-electron density with a maximum predicted at around $8 \cdot 10^{18}~cm^{-3}$.

The second term on the right side of equation 1 accounts for the contribution of IMP. A way to account for a wavelength dependence of this coefficient can be to rely on the semi-classical Drude formalism, where the avalanche coefficient is expressed as,

$$\alpha = \frac{\sigma}{BG} = \frac{1}{BG}\frac{e^2}{n(\omega)\cdot m^* c\, \epsilon_0}\frac{\tau_c}{1+\omega^2\cdot\tau_c^2}, \quad \text{Eq. 2}$$

Where $\sigma$ is the inverse Bremsstrahlung absorption cross section, $n(\omega)$ the refractive index of the irradiated material at the excitation wavelength, $c$ the speed of light, $\epsilon_0$ the vacuum permittivity and $\tau_c$ the free electron scattering time. This last term depends on the free electron density and electron temperature, being a dynamic parameter during the pulse [44,45]. This dependency on the complex light-matter interaction process makes hard to derive a robust description of $\tau_c(t, n_e)$. On the framework of this work, this description would be even harder since a large spectral range and different materials should be accounted.

An interesting study of Rajeev et al. [14] revealed a field dependence of the apparent avalanche coefficient by irradiating dielectrics with different pulse durations. In our work, inspired by this study, we decided to extract from our ablation threshold measurements, apparent avalanche coefficients by fixing an absorbed energy density target to be reached at the end of the irradiation pulse for ablation. While this approach will not permit to describe the detailed aspect of the underlying physics, it leads to the interesting possibility for a phenomenological modelling of the wavelength dependence of femtosecond laser ablation using the usual electron density rate equation (Eq. 1).

## RESULTS AND DISCUSSION
### A. Fluence ablation threshold as a function of the wavelength

The measured ablation fluence thresholds as a function of the wavelength are represented in figure 4 (a) for the four different samples. Tables containing the values ($w_{0,Liu}$, $F_{th}$ and errors) and further information can be found in the supplemental material.

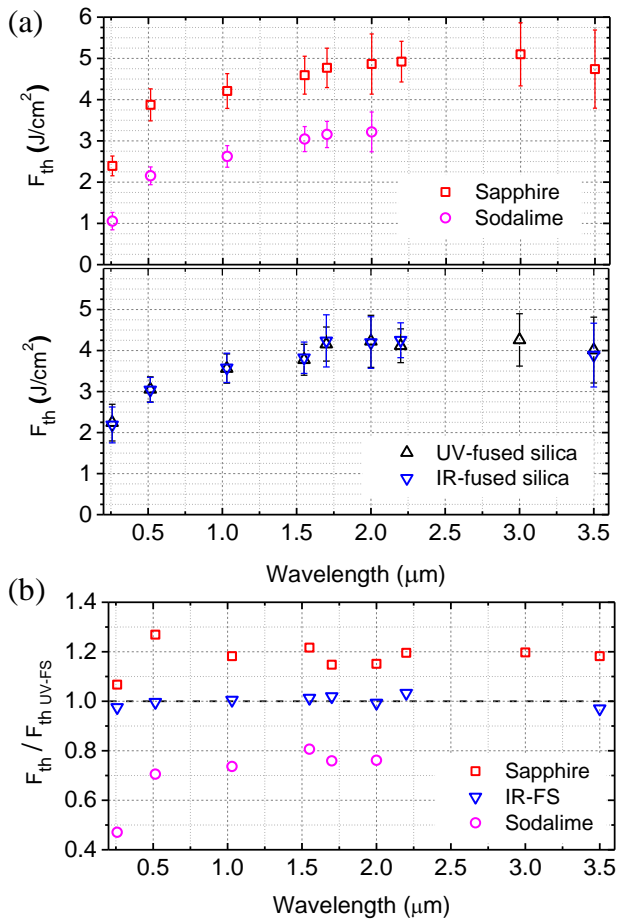

**Figure 4.** (a) Single-pulse fluence threshold for femtosecond laser ablation of dielectrics with different wavelengths. (b) Measured ablation fluence thresholds for the different materials normalized to values obtained for UV-fused silica (reference for material comparisons).

As a first general observation, one can note that despite an exploration over more than five octaves the relative changes of the fluence threshold does not exceed a factor of 3. This and the positions for the minimum (ultraviolet) and the maximum threshold values (near infrared) are common features for all the considered dielectric materials. Concentrating on fused silica, we report on a maximum relative change of only 1.9, marking a significant difference with the study of Jia et al. [24] where threshold changes exceeding a factor 3 are reported. While the pulse duration (150 fs pulse duration) and the investigated spectral range (260 nm to 1.7 μm) are comparable, one can however attribute this difference to a material incubation response. We make this comment in regard to our single shot study in comparison to Jia's work [24] considering the effect of 3 applied pulses. This is also supported by a more comparable result reported by Gallais et al. [26] where a factor of difference of approximately 2.5 was reported for single-pulse irradiations for fused silica (310 nm – 1030 nm, pulse duration approx. 100 fs).

A second important and general tendency for all materials is a near-invariance of the measured thresholds all over the infrared range (from 1.03 to 3.5 μm). After a steep increase on the fluence ablation threshold from the UV to the near-infrared (258 to 1030 nm), a very moderate increase on the short-wave infrared regime followed by a stabilization up to 3.5 μm is noted. This result reproduces the observation reported by Jia et al. up to 1.7 μm [21] in fused silica and calcium fluoride, and confirms the constancy of the ablation response deeper to the infrared (up to 3.5 μm). For an interpretation of this characteristic response, calculations and discussions about the relative role of the different ionization processes will be pursued in section B.

Looking now in more detail at the material responses, we clearly observe differences between them over all the spectral range: reaching the highest values for sapphire and the lowest for sodalime. Apart from that, two observations deserve a particular attention.

First, similar values are obtained on the two fused silica samples. As mentioned on the sample description, the linear optical response of the two samples is different, with absorption observed in the IR-fused silica at wavelengths around 250 nm (approx. 5 eV). This indicates a different optical material bandgap, which should be reflected by different nonlinear absorption response and finally different fluence thresholds for modification. Nevertheless, Figure 4 (b) does not confirm this point by showing very similar thresholds found for both materials independently of the wavelength. Accordingly, the discussion on the relative change of the fluence threshold values between materials cannot be reduced to the simple bandgap difference.

Second, the comparison between samples shown in figure 4 (b), reveals also an almost constant ratio between measured thresholds ~ × 1.2, between fused silica and sapphire. Interestingly, some previous studies irradiating those two materials with different conditions found similar relative differences. Puerto et al. [46], (120 fs, 800 nm and angle of incidence of $53^o$) reported an ablation fluence thresholds of 5.4 J/cm2 for fused silica and 7.0 J/cm$^2$ for sapphire, that is a relative difference of 1.30. Garcia-Lechuga et al. [47] (120 fs, 800 nm) reported fluence thresholds at 3.7 J/cm$^2$ for fused silica and 4.6 J/cm$^2$ for sapphire. These values and the corresponding difference factor of 1.24 compare remarkably well with our measurements for λ=1030 nm. Overall, the figure 4 (b) tends to confirm more the material-dependent differences than a wavelength-dependent response. This holds very well for all studied cases at the exception of the UV-domain.

This difference in material responses is again not directly related with the material bandgaps (sapphire: 8.8 eV and UV-fused silica: 8.9 eV), leading to a conclusion which deviates from the one extracted from the work by Gallais and coworkers [26]. Looking at the other material properties that could be associated with the observed differences between the two materials, one can refer to the findings of Grehn and coauthors [48] and compare the ratio difference between the dissociation energies of two materials:

$$\frac{E_{diss}(Sapp)}{E_{diss}(FS)} = \frac{78\ kJ/cm^3}{65\ kJ/cm^3} = 1.2 \qquad \text{Eq. 3}$$

This ratio compares very favourably with the measured fluence thresholds for ablation. Therefore, the dissociation energy could be hypothesized as the parameter of major importance for understanding the relatively modest differences between the fluence threshold for ablation observed between dielectrics. This leads to the hypothesis of nonlinear energy deposition with femtosecond pulses nearly independent of the considered dielectric material (same wavelength dependencies) and of threshold values mainly related to a critical dissociation energy to be reached and depending on materials (constant differences). This second observation appears to hold true when comparing fused silica and sapphire with pulse durations on the range of few hundreds of femtosecond. However we cannot directly extrapolate this observation to other pulse durations (e.g. Temnov et al. [17] by using 800-nm 50-fs pulses report a × 1.7 factor between measured thresholds).

Additional comparisons extended to other materials would be also needed to confirm the generality of this observation. Nevertheless, this view on referring to the dissociation energy can support the observation of identical response for the two fused silica samples. For these two materials, the different linear absorption response is attributed to the material purity level (defect absorption). According to the very deterministic ablation response in the ultrafast regime, it is commonly accepted that the nonlinear ionization response becomes independent to defects because the number of produced carriers during the pulse largely exceeds the material defect densities. Also, we expect the dissociation energy to be a macroscopic material property that will not depend on low-density defects. Overall, the identical ablation thresholds found for these samples is then interpreted as a direct consequence of these retained considerations for our analysis.

## B. Relatives roles of ionization processes at fluence threshold levels.

From now, we will concentrate only on fused silica and sapphire, being two of the most studied and widely used dielectrics for applications. For the particular case of fused silica, modelling will only consider the experimental data and bandgap of UV-fused silica.

In Figure 5, we represent the Keldysh adiabatic parameter calculated considering the peak intensity at fluence threshold conditions (values on fig. 4) for sapphire and fused silica. It can be observed on both cases that values well above unity are only occurring for 258 nm and 515 nm. Therefore, only for the UV-VIS spectral region, MPI is expected as the dominant photoionization mechanism. From 1.55-µm wavelength (NIR and MIR radiation) the Keldysh parameter becomes significantly smaller than one, which corresponds to a dominance of TI. It is striking to note that the most commonly studied ablation conditions near 1-µm actually correspond to the apparently most complex situation with a mix between the two processes (γ~1).

On a first attempt to explain the threshold tendency observed in Fig.4, we can reasonably attribute the increasing values of fluence threshold for ablation from UV to VIS to an increase on the multiphoton order and so a corresponding decrease of the MPI yield. The near invariant ablation fluence threshold values in NIR and MIR is respectively consistent with the expected wavelength-independent response when TI becomes the dominant ionization mechanism [25]. This leads to an overall qualitatively consistent picture but with a validity obviously limited due to an absence of consideration of the collision-assisted ionization processes.

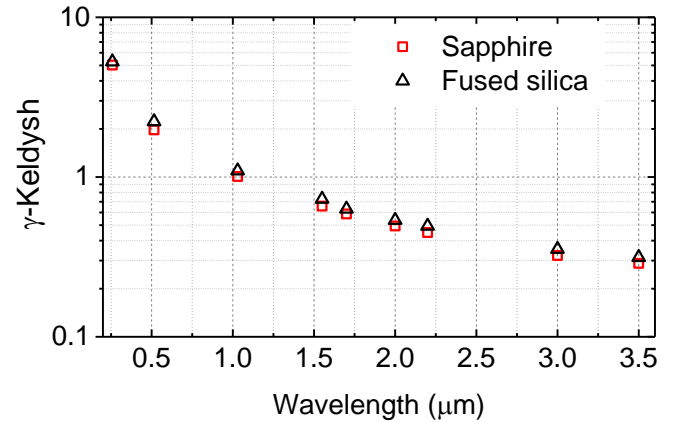

**Figure 5**. Calculated Keldysh adiabatic parameter for fused silica and sapphire as a function of the wavelength at peak intensities corresponding to the experimentally obtained fluence threshold of ablation.

In order to enlarge the information that can be extracted from experimental data, we have performed calculations of the free electrons generated by SFI for each experimentally determined threshold conditions, as shown as the example in figure 3 (b). The calculated values of $n_{e\ by\ SFI}$ for fused silica and for sapphire are plotted on figure 6 (a). Under UV radiation electron densities generated by SFI exceeds $10^{21}$ cm$^{-3}$, corresponding to high excitation values above 1% of the total valence electron density ($n_T = 1 \cdot 10^{23}$ cm$^{-3}$). However, for VIS, NIR and MIR regimes, the percentage of generated free electrons by SFI shown in figure 6 varies between 0.1% and 0.001% of the valence density. These low percentages are far below the physical conditions expected for the occurrence of ablation (e.g. 2% set by Christensen and Balling [43]), and thus directly indicates a largely increasing importance of impact ionization mechanism (IMP) in energy deposition.

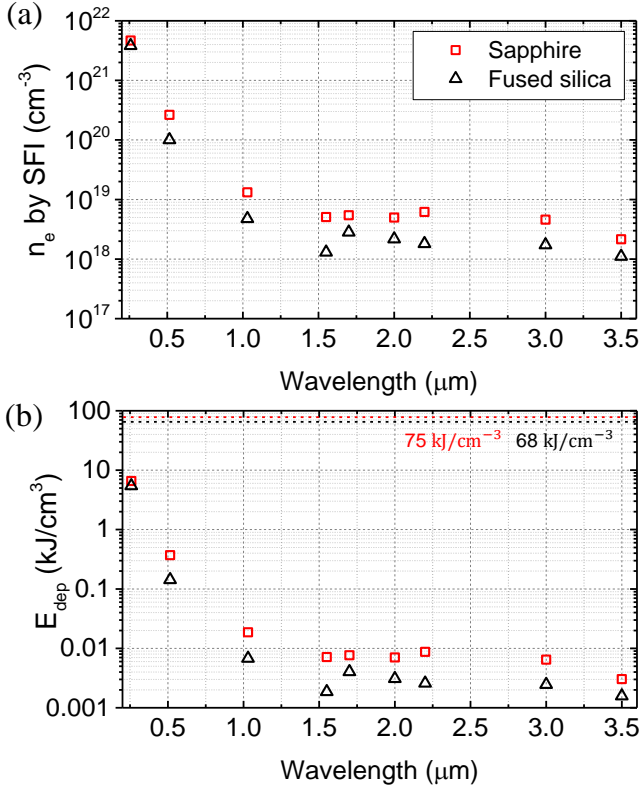

**Figure 6**. (a) Numerical calculations of the maximum free electron density obtained by SFI in UV-fused silica and in sapphire at the fluence threshold level. (b) Estimations of the corresponding minimal deposited energy to create the calculated free-carrier population. Dash lines represent the dissociation energy of both materials for comparisons.

After showing in section A a correlation between ablation fluence threshold and dissociation energy in a similar way that it was also stated by Grehn and coauthors [48], we propose in this work a simplified expression to evaluate the level of energy deposition and to establish a direct comparison with the dissociation energy. Ignoring the potential energy stored by acceleration of free electrons in the conduction (absence of hot electrons), we convert the created free-carrier density directly in deposited energy density, $E_{dep}$, according to:

$$E_{dep}[\text{kJ/cm}^{-3}] = BG\ [\text{kJ}] * n_e\ [\text{cm}^{-3}] \quad \text{Eq. 4}$$

With this conversion, we simply consider that each electron promoted from the valence to the conduction band has required the absorption of sufficient energy from the laser pulse to span the material band gap of energy *BG*. Those estimates at threshold conditions for sapphire and fused silica are plotted on Figure 6 (b). The values obtained for the full spectral range are largely below the dissociation energy of the considered materials.

In order to account for the potential contribution of IMP, we calculate the apparent avalanche coefficient values for each considered wavelength ($\alpha(\lambda)$), so that the deposited energy equals the dissociation energy. In practice, we implement a calculation procedure based on eq.1 and eq.4. which iteratively tests different avalanche coefficient values until a match is obtained between energy densities. The obtained avalanche coefficients for both materials are plotted in Figure 7. As shown in the figure, higher values are obtained for fused silica than for sapphire. This is a direct consequence of a deposited energy by SFI which is less efficient than for sapphire (figure 6 (b)) while dissociation energies are not differing much. Additionally, both materials show a similar tendency with low variations in NIR-MIR spectral range and an increase of typically a factor 2 from the shortest considered wavelength in UV to the end of the VIS region.

Therefore, observing the tendencies on Figure 6 (a) and Figure 7, we conclude on nearly invariant SFI and IMP contributions in the NIR-MIR for the studied dielectric materials. It is worth noting that this picture differs from the interpretation of Jia and co-workers [24]. A plateau on the ablation fluence threshold in the infrared was also observed in their study but this trend was interpreted based on an increasing importance of IMP compensating an expected decrease of the SFI yield towards the infrared.

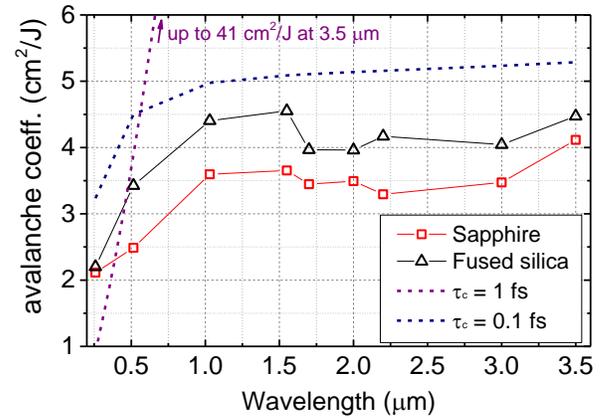

**Figure 7**. (Symbols) Estimated wavelength-dependent avalanche coefficients for deposition of an energy density equal to the dissociation energy of the material (eq.1 and 5) at fluence threshold conditions for ablation (measured). (Dashed lines) Calculated avalanche coefficients following the Drude model (eq. 2) for fused silica (considering two different collision rates).

As stated before, a simple way to calculate avalanche coefficients without the presented comparison between simulation and measurement results is to use the Drude model (eq 2). For comparison, we present in figure 7 (dashed lines) calculations taking the electron mass for $m^*$, the physical properties of fused silica (band gap and refractive index [49]) and two different values for the carrier scattering time ($\tau_c = 1$ fs and $\tau_c = 0.1$ fs). In comparison to our analysis, the Drude model calculations with $\tau_c = 1$ fs lead to values that underestimate the contribution of IMP in the UV and strongly overestimate it at NIR and MIR. However, the calculation $\tau_c = 0.1$ fs reproduces much better the general trend even if a constant overestimation is observed all over the spectrum. As reported in works exploring femtosecond excitation dynamics of dielectrics, $\tau_c$ is not rigorously a constant and is expected to depend on the free-electron density among other parameters. For instance, simulations on the reflectivity response of sapphire [47] (pulses of 120 fs at 800 nm) reveal a scattering time (inverse of the scattering rate showed on the reference) varying from 1 fs at the beginning of the pulse to values under 0.1 fs from the time at peak intensity for material modification conditions. Accordingly, the relatively good qualitative correspondence obtained with an "averaged" fixed value of 0.1 fs can appear reasonable.

For quantitative analyses, it remains important to note the significantly different avalanche coefficients derived from our simulations and measurements in comparison to those that can be predicted using the Drude model. While the underlying physical processes behind the determined wavelength-dependence of the avalanche coefficient will require more investigations, the obtained data can already serve (with extrapolation) to improve the predictive precision of models based on the standard electron density rate equation. This holds for instance for recent modelling efforts considering a modified electron density rate equation accounting for the potential influence of ultrafast carrier trapping processes (modified version of Eq. 1). The aim of such developments is to assess the expected specificities of the modification response of dielectric materials with the advent of new ultrafast MIR laser technologies [50].

**C. Crater topography as a function of the wavelength**
In previous sections, we have concentrated on the measured ablation fluence threshold for the analyses. However, more experimental observables become available above threshold conditions to enrich the discussion on the wavelength-dependence of the material responses. In particular, we focus here on the changes of the superficial and volumetric aspects of the produced modifications.

As reported in a previous publication [39] but with a data set limited to two wavelengths (515 nm and 1550 nm), we confirm here (fig. 2 and Supplemental Figure S1) that sapphire exhibits always neat craters, without noticeable elevations on the crater borders. In contrary, as can be seen on fig. 2 (a) ($\lambda = 1.7$ μm), sodalime glass presents notorious elevations at the edges, which has been associated previously to the flow and resolidification of molten material [51]. This observation in sodalime glass independent on the laser wavelengths (see also Supplemental Figure S2) together by similar observations on fused silica samples (Supplemental Figure S3) confirm that the superficial morphology of the modifications, and therefore the quality of the laser machining depends more on the material characteristics than the applied laser wavelength. Indirectly, this also confirms that the drastic changes of contributions from SFI (MPI or TI) and IMP expected from UV to MIR do not have a clear influence on the processing quality.

Another interesting observable in this context is the obtained crater depths revealing the ability of the different wavelengths to penetrate in the excited transparent materials. This provides both access to more understanding of the physics of interaction and to the ablation efficiency of interest for the potential industrial applications. For clarity of the presentation, we present and discuss here the results obtained on sapphire and provide similar data sets obtained for the other dielectrics (Supplemental Figures S4 and S5). In figure 8 (b-c) we report the maximum crater depth found in craters (at centre) produced in sapphire as a function of the irradiation peak fluence ($F_0$). For the ease of comparisons between wavelengths, the x-axis is represented normalized with respect to the fluence threshold of ablation ($F_0/F_{th}$).

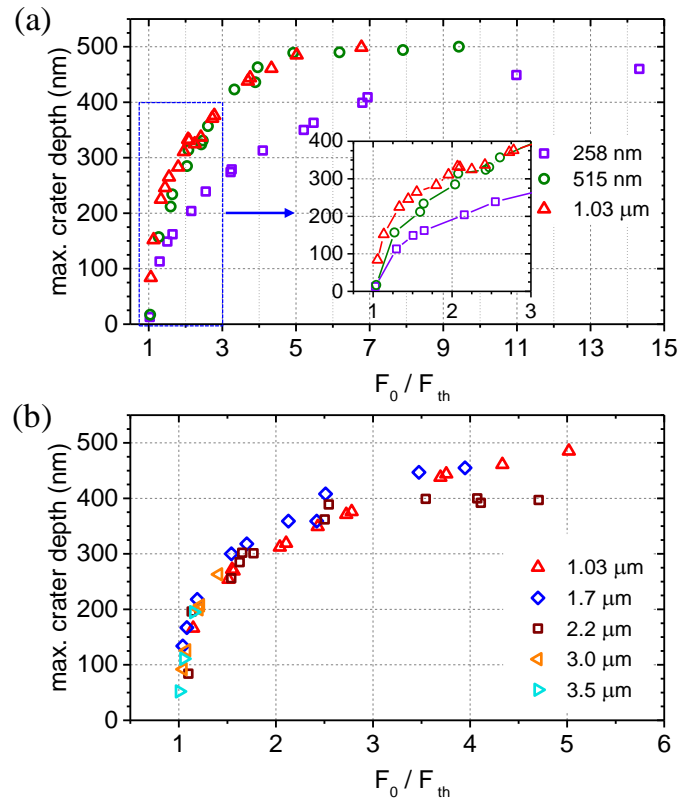

**Figure 8**. (a-b) Maximum crater depth produced sapphire under a single-shot irradiation at different excitation levels ($F_0/F_{th}$). The inset in (a) shows a zoom of the data close to the fluence threshold.

A first observation is a systematic saturation behaviour on the achievable ablation depth. It is also striking to note a near identical saturation depth for all the considered wavelength at ≈ 500 nm here for sapphire (≈ 450 nm for Fused silica, shown in Supplemental Figures S4). Despite the good precision in focus positioning (< 10 μm that is significantly smaller than the confocal parameters of all beams), the small differences of the observed plateau levels can be attributed to this question. We make this comment after verifying the sensitivity of this measurement to relative changes of the focusing depth with micrometre precision (Supplementary Note 2).

Interestingly, this measured invariance of maximum ablation depth deviates significantly from the predictions of the modelling work of Christensen and Balling [43] where the change of the apparent nonlinearity of interaction by varying the material bandgap (3 to 9 eV) or the wavelength in a spectral range much smaller than the one studied here (between 400 and 1060 nm) was leading to more significant variations (deeper machining at longer wavelength). While such experimental quantities are not accessible by our current model which does not describe pulse propagation into the excited dielectrics, this shows a possibility for improved and more robust simulations by implementing the wavelength-dependent avalanche coefficients derived from our work.

Looking at the details of the ablation depths near threshold conditions, one can note slightly different slopes as a function of the normalized fluence between 1030 nm and 515 nm and significantly reduced ablation rates at 258 nm (see inset in Fig. 8(a)). This variation of the energy dependence can be directly attributed to the dominating role of MPI (few photons) expected for UV. This causes a nonlinear Beer-Lambert law that requires more energy increase to see the ablation front penetrating deep inside the matter. Again, given the modest wavelength-dependence of nonlinear absorption in the NIR-MIR domain and the exponential growth of carrier-density based on the dominating avalanche mechanism, one can qualitatively understand the more abrupt depth rise before saturation as observed with near invariant measurements in this spectral domain (Fig. 8).

**SUMMARY AND CONCLUSIONS**

From this work aiming at a broadband study of single shot ablation of dielectrics (fused silica, sapphire and sodalime) with 200-fs pulses, we report on near-invariant responses over a domain ranging from VIS to MIR up to 3.5 μm. This holds for the measured fluence thresholds for ablation and ablation performance (crater morphologies and ablation efficiency) and complements our previous findings on the nonlinearity invariance of machining resolution due to the deterministic threshold-nature of ablation in ultrafast regimes [39,40,52].

For fundamental considerations, analyses show a change of dominant photoionization mechanism, from multiphoton to tunnelling ionization, when varying the wavelength from VIS to MIR without significant change on the observed ablation performance. We also conclude on a contribution of photoionization to energy deposition in ablation conditions that remain weak over the broadband range of investigated wavelengths. However, multiphoton and tunnelling are necessary for seeding the impact ionization mechanism that appears extremely dominant in the whole spectral range (above 99 % of the total deposited energy).

For technological considerations, these conclusions open a wide range of possibilities for processing system optimizations with the advent of new ultrafast laser technologies emitting in different spectral domains and giving access to different performance advantages. For instance, one can refer to the kW power level demonstrated with compact fiber laser technologies in the mid-IR (mainly Thulium-Based fiber lasers) [53]. This can offer solutions for high-throughput production capabilities without compromising machining qualities.

Finally, the only modest variations observed in the UV make that this spectral domain remain probably the most promising one for accessing even higher processing performances. As shown with typical two-fold decrease of the ablation thresholds, this part of the spectrum is highly favourable for efficient energy coupling to materials, but it is obviously also an optimum in dealing with diffraction limits and thus achieving high-lateral processing resolutions. Interestingly, we also show in this work a more modest sensitivity of machining depth to the applied energy and material for near threshold conditions (in comparison to VIS-NIR-MIR). This reveals a situation beneficial for reliable material removal with extreme longitudinal resolution. Recently, impressive demonstrations have been made on reliable matter removal with nanometric precision (depth) at the surface of dielectrics materials using NIR femtosecond pulses [54]. On the basis of our findings, even higher precision can be expected using similar pulses (same stability) in the UV domain.

Overall, we believe that this comprehensive report made on the wavelength-dependence of ablation is timely considering the increasing diversity of ultrafast laser technologies now emitting in various spectral domains. It is also expected to be beneficial for the development of improved simulations of the complex interaction problem. In particular, it sheds light for better understanding on the avalanche coefficient, an important physical parameter in the field of laser processing of materials.

**ACKNOWLEDGEMENTS**

This project has received funding from the European Research Council (ERC) under the European Union's Horizon 2020 research and innovation program (Grant Agreement No. 724480).

# Supplemental Material for "Wavelength-independent performance of femtosecond laser dielectric ablation spanning over three octaves"

Mario Garcia-Lechuga [1,2] *, Oliver Utéza [1], Nicolas Sanner [1], David Grojo [1†]

[1] *Aix Marseille Université, CNRS, LP3, UMR7341, 13288 Marseille, France*

[2] *Instituto de Óptica, IO-CSIC, Serrano 121, 28006 Madrid, Spain*

*mario.garcia.lechuga@csic.es

[†] david.grojo@univ-amu.fr

## Table of contents

## Table S1: Ablation fluence threshold values

|  | Fluence threshold of ablation (J/cm$^2$) | | | |
|---|---|---|---|---|
| $\lambda(nm)$ | Sapphire | UV-Fused silica | IR-fused silica | Sodalime glass |
| 258 | 2.4 ± 0.2 | 2.2 ± 0.4 | 2.2 ± 0.4 | 1.1 ± 0.2 |
| 515 | 3.9 ± 0.4 | 3.1 ± 0.3 | 3.0 ± 0.3 | 2.2 ± 0.2 |
| 1030 | 4.2 ± 0.4 | 3.6 ± 0.4 | 3.6 ± 0.4 | 2.6 ± 0.3 |
| 1550 | 4.6 ± 0.5 | 3.8 ± 0.4 | 3.8 ± 0.4 | 3.0 ± 0.3 |
| 1700 | 4.8 ± 0.5 | 4.2 ± 0.4 | 4.2 ± 0.4 | 3.2 ± 0.3 |
| 2000 | 4.9 ± 0.7 | 4.2 ± 0.6 | 4.2 ± 0.6 | 3.2 ± 0.5 |
| 2200 | 4.9 ± 0.5 | 4.1 ± 0.4 | 4.3 ± 0.4 | --- |
| 3000 | 5.1 ± 0.8 | 4.3 ± 0.6 | --- | --- |
| 3500 | 4.7 ± 0.9 | 4.0 ± 0.8 | 3.9 ± 0.8 | --- |

*Table S1.: Values reported on figure 4(a) of the article*

## Supplemental Figure S1: Details on crater morphology on sapphire

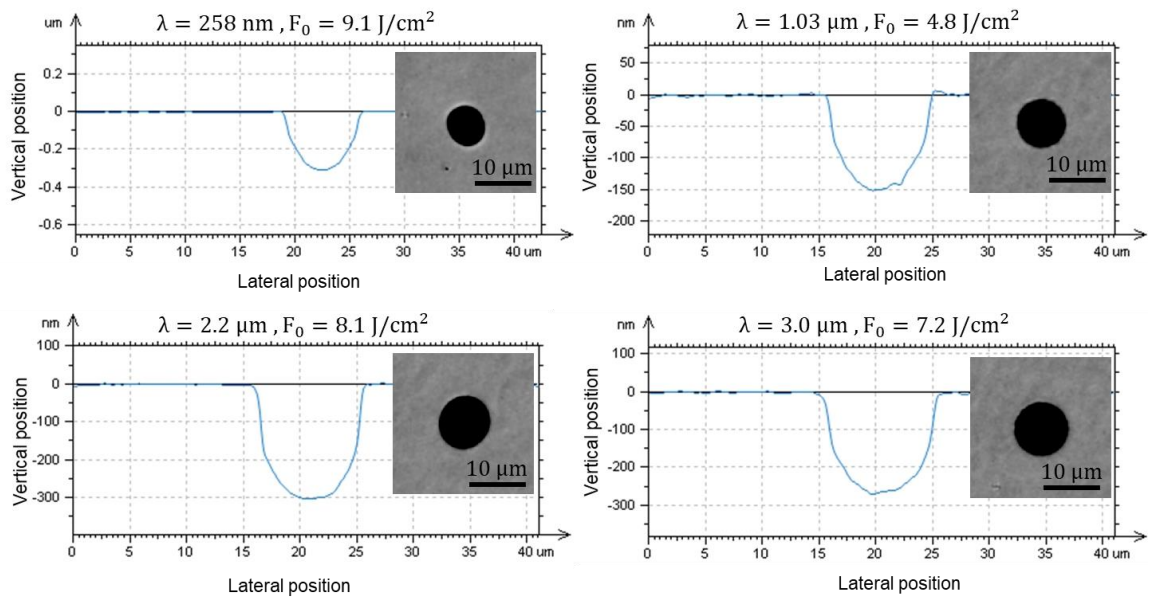

*Figure S1.: Examples of the topographies induced on sapphire after a single shot (experimental conditions indicated on the images). The vertical depth profiles correspond to the measured profile along a line (x-axis) passing at the centre of the modification shown at right.*

*Supplemental Material for "Wavelength-independent performance of femtosecond laser dielectric ablation spanning over three octaves" M. Garcia-Lechuga, O. Utéza, N. Sanner, D. Grojo*

## Supplemental Figure S2: Details on crater morphology on sodalime glass

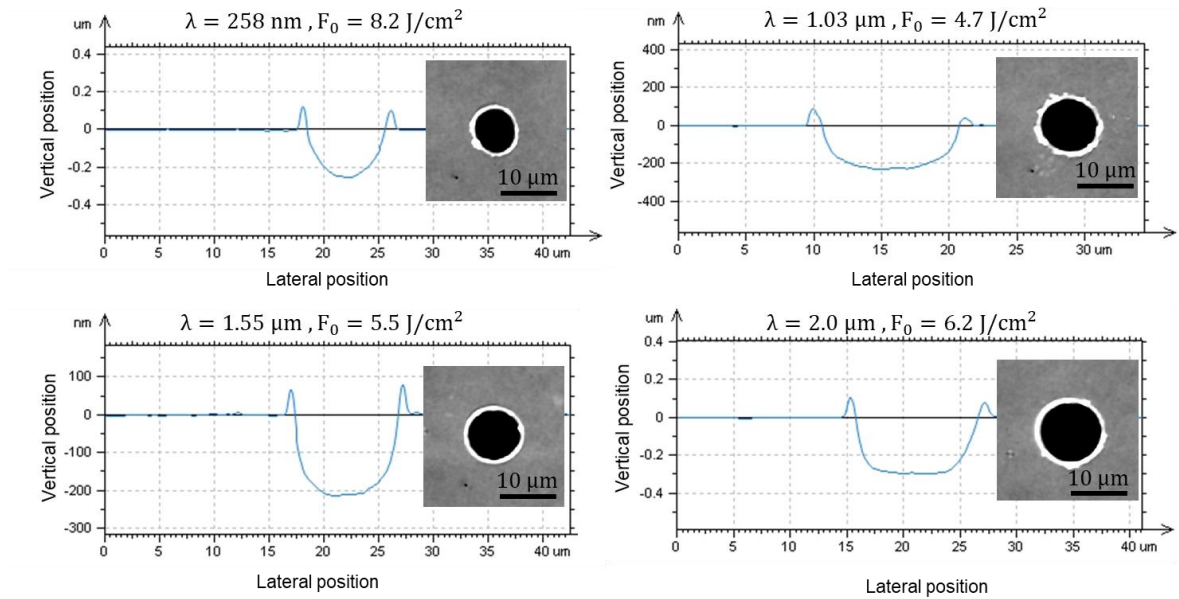

*Figure S2: Examples of the topographies induced on sodalime glass after a single shot (experimental conditions indicated on the images). The vertical depth profiles correspond to the measured profile along a line (x-axis) passing at the centre of the modification shown at right.*

## Supplemental Figure S3: Details on crater morphology on fused silica samples

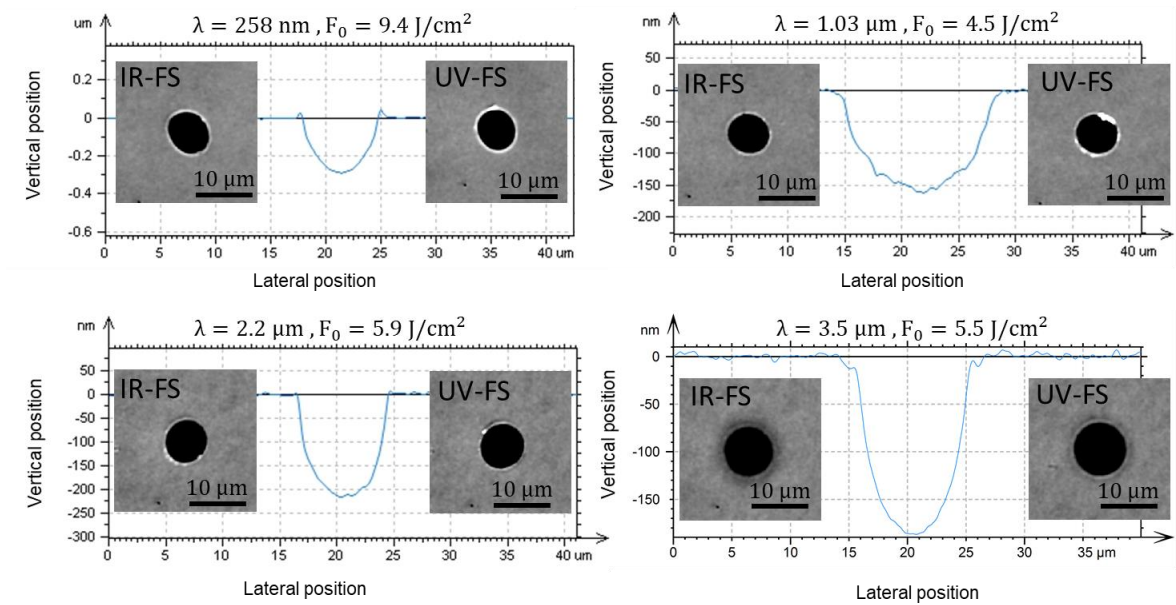

*Figure S3: Examples of the topographies induced on UV-fused silica after a single shot (experimental conditions indicated on the images). The vertical depth profiles correspond to the measured profile along a line (x-axis) passing at the centre of the modification shown at right. Additionally, at the left part of each profile crater images produces on IR-fused silica under the same conditions are shown.*

*Supplemental Material for "Wavelength-independent performance of femtosecond laser dielectric ablation spanning over three octaves" M. Garcia-Lechuga, O. Utéza, N. Sanner, D. Grojo*

**Supplemental Figure S4: Maximum crater depth versus fluence in fused silica**

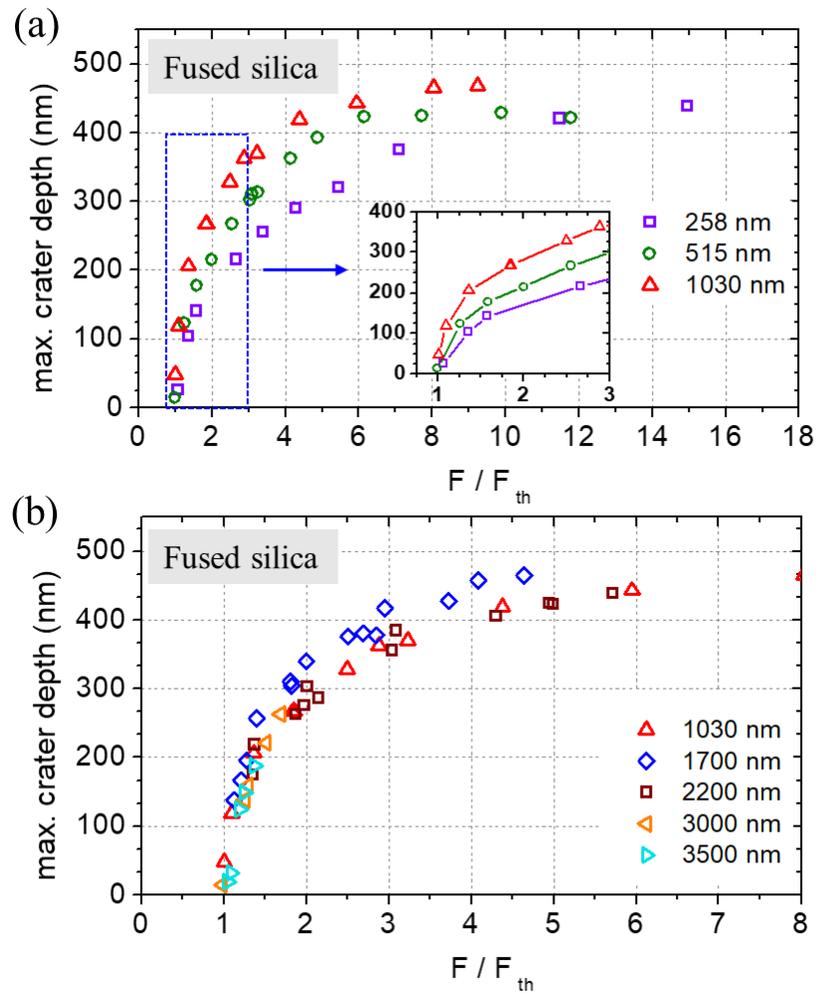

*Figure S4: (a-b) Maximum crater depth produced in UV-fused silica under a single-shot irradiation at different excitation levels ($F/F_{th}$). The inset in (a) shows a zoom of the data close to the fluence threshold*

## Supplemental Figure S5: Maximum crater depth versus fluence in sodalime glass

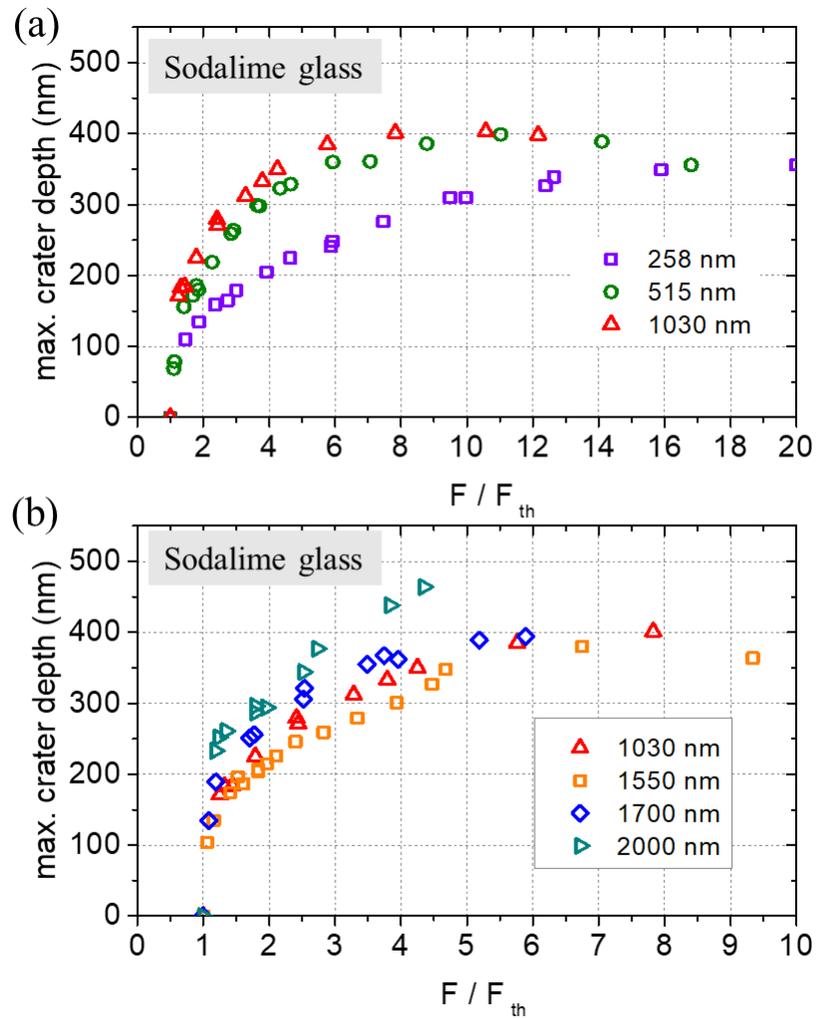

*Figure S5: (a-b) Maximum crater depth produced in sodalime glass under a single-shot irradiation at different excitation levels ($F/F_{th}$).*

## **Supplementary Note 1:** Determination criteria of ablation fluence threshold and its errors

As mentioned in the article, the fluence threshold determination is done by following the Liu's methodology [1]. This method, by definition only valid for perfect Gaussian beams, has been extended to be applicable to Airy-like beam shapes [2], resulting from the truncation of beams profiles with a circular aperture just before the focusing lens.

In our studies, beam truncation was not applied (power transfer of the aperture $P_T = 100\%$) for 258 nm and 515 nm considering that beam profiles are Gaussian after harmonic generation (see for instance the beam profile at 515 nm in ref [3]). For the other wavelengths, as shown by beam imaging for 1030 nm and 1550 in ref. [2], the deviation from a perfectly Gaussian beam makes important the beam truncation for a correct determination of the fluence threshold of ablation. Beam truncation details for each wavelength can be found on the table at the end of this note (Table S2).

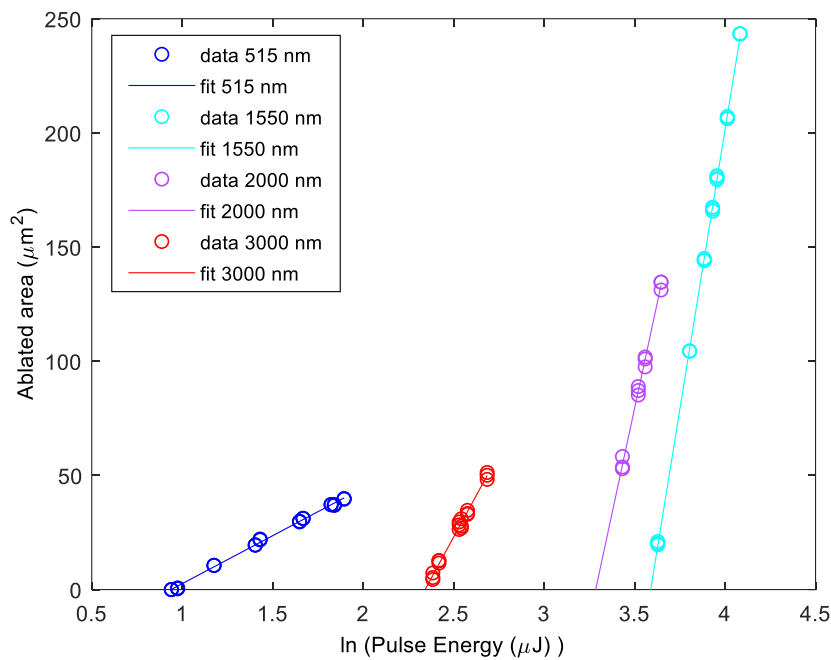

*Figure S6: Representation following the Liu's method (crater area versus the pulse energy in logarithmic scale) of modifications produced at the surface of a sapphire sample with a single laser pulse. The different data corresponds to different wavelengths used (see legend).*

In Figure S6, some examples (wavelengths 515 nm, 1550 nm, 2000 nm and 3000 nm) of the linear fitting procedure following Liu's method for modifications produced in sapphire are shown. The equivalent Gaussian beam waist (radius) retrieved by the linear fitting for the different wavelengths (those represented on the figure and the others) are presented in Table S2. Additionally, on the table are shown the maximum excitation energy (*M.E.*) used for the linear fitting procedure, that together with the $P_T$ are important values to determine the correction factor to be applied for the fluence threshold determination when applying the Liu's method to Airy-like beams [2].

Regarding the estimated error for fluence threshold determination, as we indicated on Ref. [2] the Liu's methodology for Airy-like beams should incorporate an error not less than 5%. On this work, in order to be more conservative, we provide values with a 10% of error, with higher values for 2000 nm, 3000 nm, 3500 nm and 258 nm. For 2000 nm and 3000 nm, the estimated error is >10 % because of the limited energy values above the threshold ($<1.5 \times E_{th}$) which reduces the accuracy of a linear fitting. For 3500 nm, together with the previous mentioned error, the need of using a higher power transfer (less truncated beam and as consequence, less Airy-like beam shape) makes the estimated error also >10 %. . Finally, for 258 nm, the absolute value of the error bar on the ablation fluence threshold is lower (0.2 J/cm²). For the two samples of fused silica, it was determined slightly higher (20% of error on the threshold) because of demanding experimental arrangement and conditions during those experiments

| $\lambda(nm)$ | $f(mm)$ | $P_T$ | M.E. | $w_{0.Liu}$ ($\mu m$) |
|---|---|---|---|---|
| **258** | 50 | 100 % | $2.5 \times E_{th}$ | 4.4 |
| **515** | 50 | 100 % | $2.6 \times E_{th}$ | 6.5 |
| **1030** | 50 | 75 % | $2.4 \times E_{th}$ | 18.1 |
| **1550** | 50 | 75 % | $1.6 \times E_{th}$ | 22.1 |
| **1700** | 50 | 75 % | $1.8 \times E_{th}$ | 20.2 |
| **2000** | 50 | 75 % | $1.4 \times E_{th}$ | 19.2 |
| **2200** | 25 | 75 % | $2.4 \times E_{th}$ | 9.6 |
| **3000** | 25 | 75 % | $1.4 \times E_{th}$ | 12.1 |
| **3500** | 25 | 85 % | $1.2 \times E_{th}$ | 14.8 |

*Table S2:* Equivalent Gaussian beam waist ($w_{0.Liu}$) obtained for each wavelength when applying Liu's method [1] for the modifications produced in sapphire. In order to apply the corrections factors [2] accounting the Airy-like beam shape of circularly truncated beams on the table are represented the experimental conditions used: focusing lens ($f$), power transfer ($P_T$) and maximum excitation energy (*M. E.*) considered for the linear fitting procedure.

## Supplementary Note 2: Maximum crater depths as a function of the sample positioning.

As indicated on the manuscript, despite the good precision in focus positioning (< 10μm) the small differences on the observed common tendency on figure 8 for 515 nm to 3.5 μm can be attributed to slight deviation from perfect sample positioning at the focus.

The different maximum crater depth values when focusing at the best focus position (z = 0 μm) and at two different positions 100 μm far from it are illustrated in figure S7

A diminution of the ablated depth when increasing the fluence is observed when placing the sample just after the focus, which can be attributed to beam perturbation (energy loses, divergence, spectral modifications) due to air ionization [4]. On the opposite direction, it is interesting to see that deeper craters (around 10% more) can be achieved when positioning the sample before the focus.

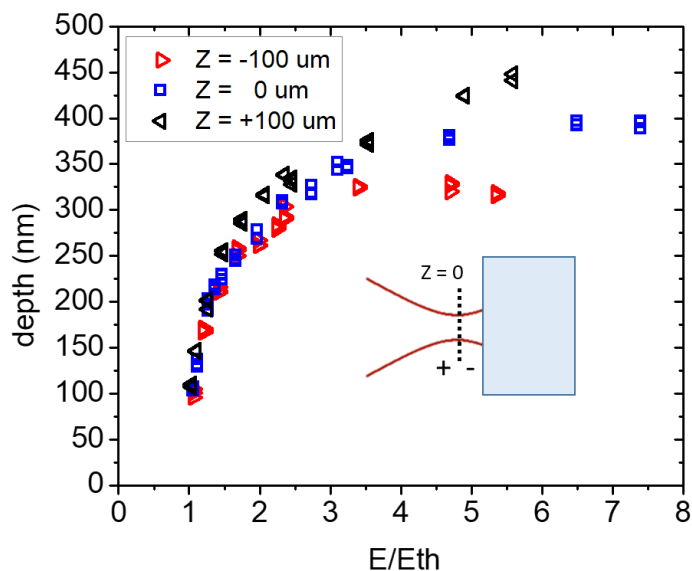

*Figure S7: Modifications on sapphire with a single pulse at 1.55 μm positioning the sample at the best focus position (z = 0 μm) and 100 μm far from it (before and after the focus). The sketch indicates the positive and negative criteria.*